\documentclass{article}
\usepackage{graphicx} 
\usepackage[final]{neurips}
\usepackage{booktabs}
\usepackage{listings}
\usepackage{hyperref}
\usepackage{caption}

\title{Better speech synthesis through scaling}
\author{James Betker}
\date{July 2022}

\begin{document}

\maketitle
\begin{abstract}
In recent years, the field of image generation has been revolutionized by the application of autoregressive transformers and DDPMs. These approaches model the process of image generation as a step-wise probabilistic processes and leverage large amounts of compute and data to learn the image distribution.

This methodology of improving performance need not be confined to images. This paper describes a way to apply advances in the image generative domain to speech synthesis. The result is TorToise - an expressive, multi-voice text-to-speech system.

All model code and trained weights have been open-sourced at \href{https://github.com/neonbjb/tortoise-tts}{https://github.com/neonbjb/tortoise-tts}.
\end{abstract}

\section{Background}
\subsection{Text-to-speech}
The field of text-to-speech (TTS) research has been largely constrained to the development of efficient models trained on relatively small datasets. This choice has been driven by:
\begin{enumerate}
    \item The desire to build efficient speech generation models that can be deployed at scale and thus must have a high sampling rate.
    \item The unavailability of very large, transcribed speech datasets.
    \item Challenges scaling the encoder-decoder model architectures traditionally used in TTS.
\end{enumerate}

\subsubsection{Neural MEL Inverters}
Most modern text-to-speech systems operate on speech data that is encoded as a MEL spectrogram. There are many compelling reasons to operate in this encoding space, but for neural networks, the most compelling reason is that it is highly spatially compressed. The MEL configuration used by the Tacotron, for example, operates at 256x compression over raw audio waveform data sampled at 22kHz, but contains most of the information found in that data.

Because of this, an entire body of research has been dedicated to finding high-quality ways to decode MEL spectrograms back into audio waveforms. A synthesizer that performs this task is generally called a “vocoder”, but I more generally refer to it as a “MEL inverter” in this paper.

Modern MEL inverters built on neural networks are incredibly sophisticated. They produce waveforms that are nearly indistinguishable from recorded waveforms to human ears, and they are highly generalizable outside of their training set. I capitalize on this work by using an implementation of Univnet\citep{univnet} as a final stage for my text-to-speech system.
\subsection{Image generation}
While TTS systems largely focus on latency, this has not been the case in other domains. For example, with image generation, more focus has been applied to training models that generate high-quality results, regardless of the sampling time. For the purposes of this paper, I dive into two bodies of research:
\subsubsection{DALL-E}
DALL-E\citep{dalle} showed how an autoregressive decoder can be applied to text-to-image generation. This is particularly appealing because of the vast quantity of research that has been poured into scaling decoder-only models in the NLP domain. 

Two important problems persist with DALL-E: first, it relies on full-sequence self-attention, which carries a cost of $O(N^2)$ compute and memory, where N is the sequence length. This is particularly troublesome when dealing with modalities like images or audio, which have large sequence lengths when dealt with naively.

Second, traditional autoregressive approaches require operating in the discrete domain. Images are encoded into sequences of discrete tokens using a quantizing autoencoder. DALL-E then models these sequences of tokens using an autoregressive prior model. This is a strength of DALL-E in terms of expressiveness, but it comes at the cost of requiring a decoder which can convert these image tokens back into the pixel values that actually comprise an image. It is my opinion that learned VQVAE decoder used by DALL-E is principally responsible for the blurry incoherence exhibited by most of it’s samples.

\subsubsection{DDPMs}
The generative model space has long been plagued by models that either exhibit mean-seeking behavior (resulting in blurriness) or mode-collapse (resulting in a lack of diversity or generalization).

Denoising diffusion probabilistic models (DDPMs\citep{ho2020denoising}) have recently arisen as the first type of generative model capable of producing crisp, coherence and diverse images. These models have been shown to be quite effective at using low-quality guidance signals to reconstruct the high-dimensional space that those guidance signals were derived from. Put another way, they are great at super-resolution.

There are two important caveats to DDPMS:
\begin{enumerate}
    \item Traditional approaches to DDPMs rely on fixed output shapes that are known before sampling begins. As a concrete example relevant to this paper, DDPMs cannot learn to convert text into audio signals because they cannot solve the implicit alignment problem between text and audio.
    \item DDPMs must be sampled from over multiple iterations. This sampling process consumes a great deal of compute, and means sampling from a DDPM will always incur a significant latency cost.
\end{enumerate}

\subsubsection{Re-ranking}
DALL-E introduced the process of “re-ranking” the outputs of autoregressive models. This process samples randomly from the autoregressive model and picks the highest quality output of k outputs for downstream use.

Such a procedure requires a strong discriminator: a model that can tell good text/image pairings from bad. DALL-E used CLIP\citep{clip}, a model trained with a contrastive text and image pairing objective.

\section{Methods}

\begin{figure}[t]
    \begin{center}
    \includegraphics[width=\textwidth]{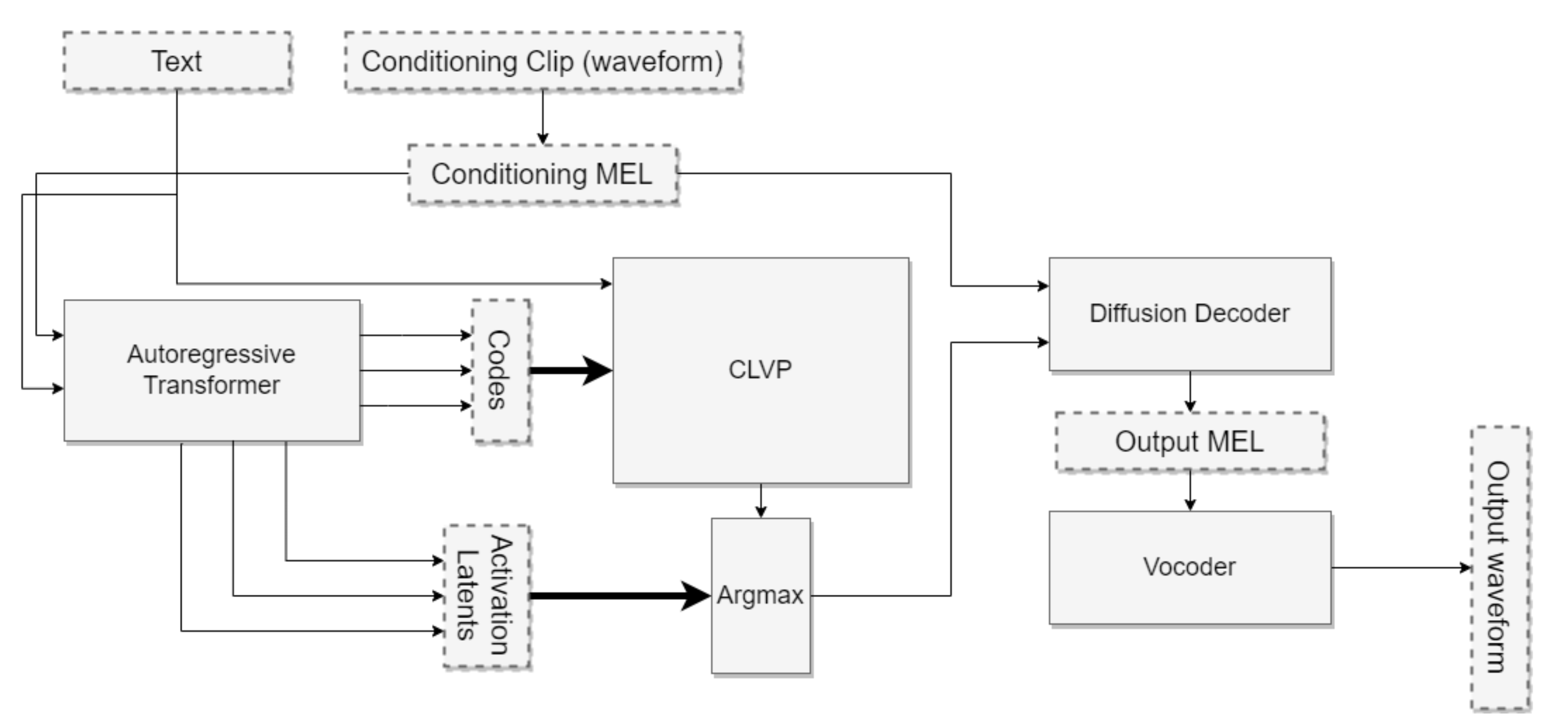}
    \end{center}
    \vskip -0.1in
    \caption{TorToise-v2 architectural design diagram. Inputs of text and a reference audio clip (for speaker cloning) flow through a series of decoding and filtering networks to produce high-quality speech.}
    \vskip -0.1in
\end{figure}

\subsection{Joining Autoregressive Decoders and DDPMs}
To review some of the conclusions drawn above:
\begin{enumerate}
    \item Autoregressive models are strong at converting between unaligned domains like vision, text and speech.
    \item DDPMs operate in the continuous domain which allows them to model expressive modalities.
\end{enumerate}

Both types of models have demonstrated the ability to scale performance with additional compute and data.

It becomes evident that when posed with a problem like generating continuous data like speech spectrograms or images, a marriage of these two approaches might have some distinct advantages.

Specifically, in inference, the autoregressive model will be used to convert a sequence of text tokens to a sequence of tokens representing the output space (in our case, speech tokens). The DDPM will then be used to decode these tokens into a high quality representation of speech.

\subsection{Applying Autoregression+DDPMs to TTS}
To build out the previously proposed system, we need to train the following neural networks:
\begin{enumerate}
    \item An autoregressive decoder which predicts a probability distribution for speech tokens, conditioned on text.
    \item A contrastive model similar to CLIP which is used to rank outputs of the autoregressive decoder.
    \item A DDPM which can convert speech tokens back into speech spectrograms.
\end{enumerate}

The architectures and training process for all of these networks largely follow the procedures found in their respective literature. Details can be found in \ref{sec:appendix}

\subsubsection{Conditioning Input}
A unique design choice made with TorToise is an additional input which is provided to both the autoregressive generator and the DDPM, which I term the speech conditioning input. 

The speech conditioning input starts as one or more audio clips of the same speaker as the target. These clips are converted to MEL spectrograms and fed through an encoder consisting of a stack of self-attention layers. The autoregressive generator and the DDPM have their own conditioning encoders, both of which are learned alongside their respective networks.

The output of these layers is averaged to produce a single vector. The vectors from all of the encoded conditioning clips are then averaged again before being fed as an input into the autoregressive or conditioning networks.

The intuition behind the conditioning input is that it provides a way for the models to infer vocal characteristics like tone and prosody such that the search space of possible speech outputs corresponding to a given textual input is greatly reduced.

\subsubsection{The “TorToise Trick”}
For the majority of the training procedure, the DDPM is trained to convert discrete speech codes into MEL spectrograms. After this process has converged, I fine-tune the DDPM on the autoregressive latent space which is pulled from the AR model outputs instead of the speech codes. This is described in detail in \ref{sec:appendix}.

The logic here is that the AR latent space is far more semantically rich than discrete tokens. By fine-tuning on this latent space, we improve the efficiency of the downstream diffusion model. I liken this to recent work showing that training decoder models conditioned on frozen text encoders to produce large efficiency gains. This fine-tuning is one of the greatest contributors to model output quality of any of the tweaks I made to the various model training processes.

\subsection{CLVP}
As mentioned earlier, a good strategy for gathering expressive outputs from generative models is using a qualitative discriminator to re-rank several outputs, then choosing only the best. DALL-E uses CLIP for this.

This same type of approach used for CLIP can be applied to speech: after all, most TTS datasets are simply pairings of audio clips and text. By training a model on these pairs in a contrastive setting, the model becomes a good discriminator for speech. 

For TorToise, I train the Contrastive Language-Voice Pretrained Transformer, or CLVP. It has many of the same properties of CLIP, but notably serves as a scoring model for use in re-ranking TTS outputs from the AR model.

To make this work efficiently in inference, I trained CLVP to pair discretized speech tokens with text tokens. This way, CLVP can rerank multiple AR outputs without the expensive diffusion model being invoked.
\section{Training}
These models were trained on a small cluster of 8 NVIDIA RTX-3090s over the period of ~ 1 year.

Specifics on how these models are trained can be found in \ref{sec:appendix}.
\section{Inference Process}
Once the four models of the framework are fully trained, the inference procedure is as follows:
\begin{enumerate}
    \item Feed the conditioning inputs and the text into the autoregressive model and decode a large number of output candidates.
    \item Use CLVP to produce correlation scores between each speech candidate and text.
    \item Choose the top k speech candidates, and for each candidate:
    \item Decode to a MEL spectrogram using the DDPM.
    \item Convert to a waveform using a conventional vocoder.
    \item When decoding the autoregressive model, nucleus sampling is used with P=.8, repetition penalty=2 and softmax temperature=.8.
\end{enumerate}

Sampling from DDPMs is a highly studied and rapidly changing field. At the time TorToise was designed, I found the sampling configuration with the best balance between quality and inference speed to be as follows:
\begin{enumerate}
    \item Algorithm: DDIM\citep{ddim}
    \item Schedule: Linear
    \item Sampling steps: 64
    \item Conditioning-Free Guidance constant: 2
\end{enumerate}

\section{The Dataset}
Since my goal was to train what is essentially a large language model, I needed a lot of data. I started with the LibriTTS\citep{libritts} and HiFiTTS\citep{hifitts} datasets, which combined contain ~896 hours of transcribed speech. I built an additional, “extended” dataset of 49,000 hours of speech audio from audiobooks and podcasts scraped from the internet. Details on how this dataset was built are in appendix I. The official LibriTTS test split was used for validation purposes.

\section{Experiments}
Text to speech systems are challenging to experimentally compare because many state of the art systems are closed source with few samples to compare against. To this end, I built my own evaluation suite which uses CLVP to produce a distance metric between real samples and generated samples, similar to the FID score used by images. I also use an open source wav2vec model to characterize the “intelligibility” of a speech segment. I have open sourced this work here. 

Past this, comparisons between the samples generated from TorToise and those generated by other papers can be found here.

\section{Conclusion}
TorToise is the latest in a line of recent state-of-the-art breakthroughs that use general model architectures. Almost no part of TorToise was designed specifically for audio processing, yet it outperforms all previous TTS models in realism. It does this by:
Embracing generalist architectures like stacks of transformer layers.
Leveraging a large, high-quality dataset.
Training at large-ish scale and high batch size.

My main take-away from this project is how incredibly strong the results are from adhering to the above 3 points. It seems likely to me that any digitized modality is subject to generative modeling using this framework.

\bibliographystyle{apalike}
\bibliography{references}

\appendix

\section{Extended Dataset Collection}
I independently built an extended TTS dataset composed of audiobooks and podcasts scraped from the web. This data was split on 500ms silences, and any audio clip between 5-20 seconds was kept. I then fed the resulting clips through a pipeline of classifiers that I trained which remove any audio with background noise, music, poor quality (such as phone calls), multiple voices speaking at once and reverb. Due to disk space limitations, I was forced to limit the amount of scraping. The end result was 49,000 hours of cleaned audio clips.

I transcribed this dataset using a wav2vec2-large model. I personally fine-tuned this model to predict punctuation, as quotation marks, commas and exclamation marks are important for the purposes of generating speech but are not generally included in the training of speech recognition models. Fine-tuning was performed on LibriTTS and HiFiTTS and the pretrained model weights and transcription scripts can be found here.

\section{Training and Architecture Details}
\label{sec:appendix}

\subsection{VQVAE}
The VQVAE used with TorToise is most similar to that of the original VQVAE by van der Oord et al. It operates on MEL spectrograms. It consists of a small residual convolutional network that compresses the spectrogram an additional 4x and produces a codebook consisting of 8192 tokens.

When training the VQVAE, I found that larger batch sizes decrease reconstruction losses, and thus used a very large batch size for my infrastructure. Input samples were constricted to 40960 PCM readings, or ~2 seconds of audio. The primary bottleneck for training the VQVAE was the dataloader.

\newpage
\begin{figure}[h]
    \begin{center}
    \includegraphics[width=\textwidth]{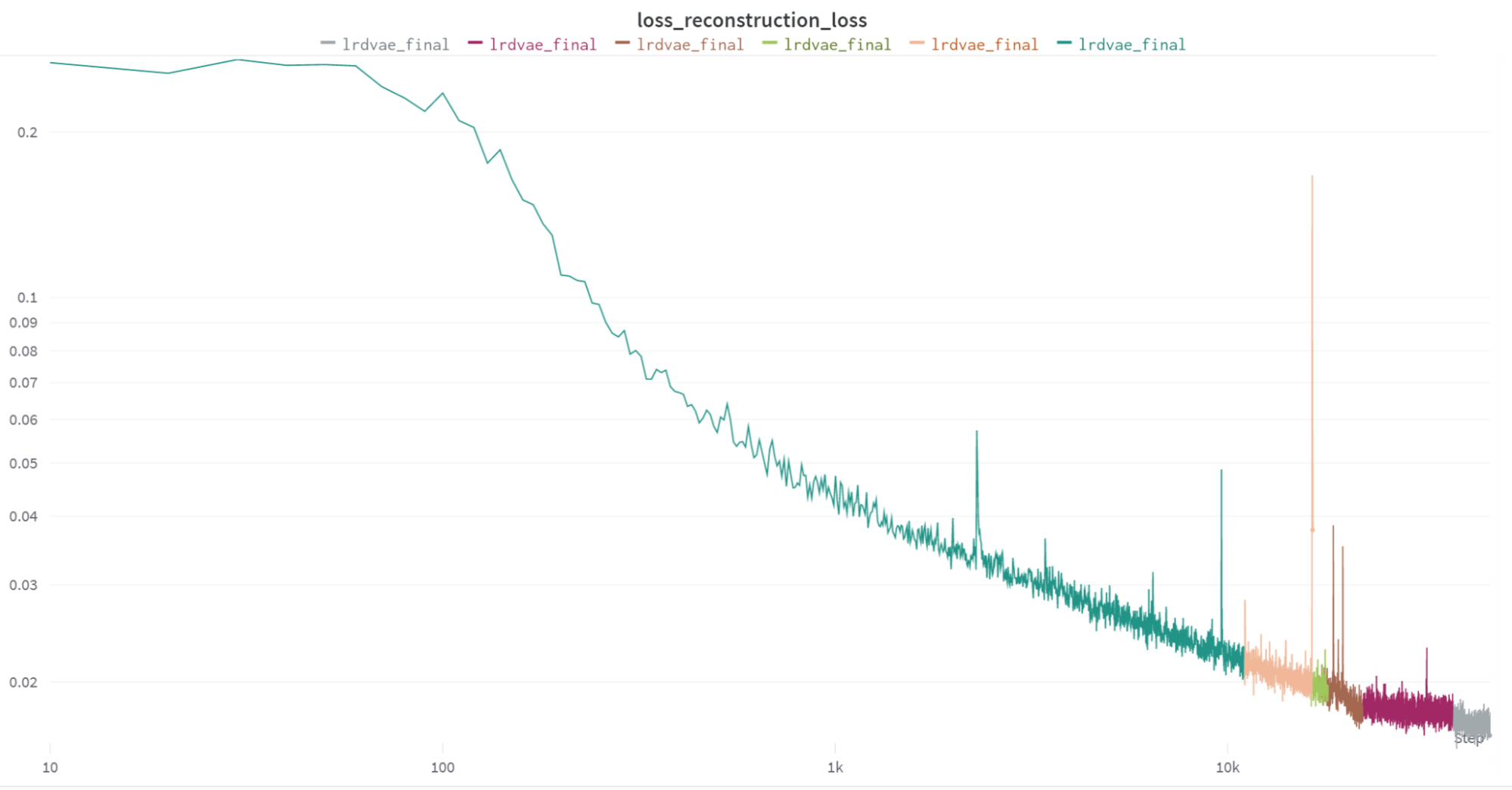}
    \end{center}
    \vskip -0.1in
    \caption{Training curves for VQVAE. Y-axis is MSE loss in log-log scale. X-axis is number of training steps.}
    \vskip -0.1in
\end{figure}
\begin{table}[h]
    \setlength\tabcolsep{4pt}
    \begin{small}
    \begin{center}
    \begin{tabular}{ll}
    \toprule
        Model shape & 1D Conv resnet, encoder + decoder \\
        Top dim & 512 \\
        Bottom dim & 1024 \\
        Codebook dim & 256 \\
        Quantizer token count & 8192 \\
        Quantization algorithm & Clustering a la original VQVAE, no restart \\
        Batch size & 8192 \\
        Total training & 360M samples \\
        Losses & MSE reconstruction loss, commitment loss \\
        LR & 3e-4 \\
        B1, B2 & .9 .9999 \\
        Weight decay & .01 \\
        EMA weights replaces LR decay with rate & .999 \\
    \bottomrule
    \end{tabular}
    \end{center}
    \end{small}
    \caption{VQVAE model details \& hyperparameters}
    \vskip -.2in
\end{table}

\newpage
\subsection{Autoregressive Prior}
The AR decoder uses a bog-standard GPT-2 architecture and generally follows the training instructions from the DALLE-1 paper. Unlike DALL-E, only dense self-attention is used. The prompt is assembled as follows:

\begin{lstlisting}
<SC><BT><T><T><T>..<T><ET><BM><M><M><M>...<EM>
SC=Speech conditioning encoding
BT=Begin text token
T=Text tokens
ET=End text token
BM=Begin MEL token
M=MEL tokens
EM=End MEL token
\end{lstlisting}

Speech conditioning encodings are learned by a separate encoder that takes in the MEL spectrogram of a related clip (another clip of the same person speaking) and produces a single vector embedding that is placed at the front of the attention context. Two encodings were produced for each training sample, which are averaged together. The maximum input length to the conditioning encoder is 132,300 samples, or 6 seconds of audio.

Learned positional embeddings are used. The MEL tokens and the text tokens get their own positional parameters. Text inputs are unpadded, MEL tokens are right padded to conform the sequence length of each batch. The maximum sequence length is 402 text tokens + 604 MEL tokens. For efficiency reasons, in the first half of training, the model only saw <6 second audio clips. After this, audio clips up to the full length (~27 seconds) were seen.

\newpage
\begin{figure}[h]
    \begin{center}
    \includegraphics[width=\textwidth]{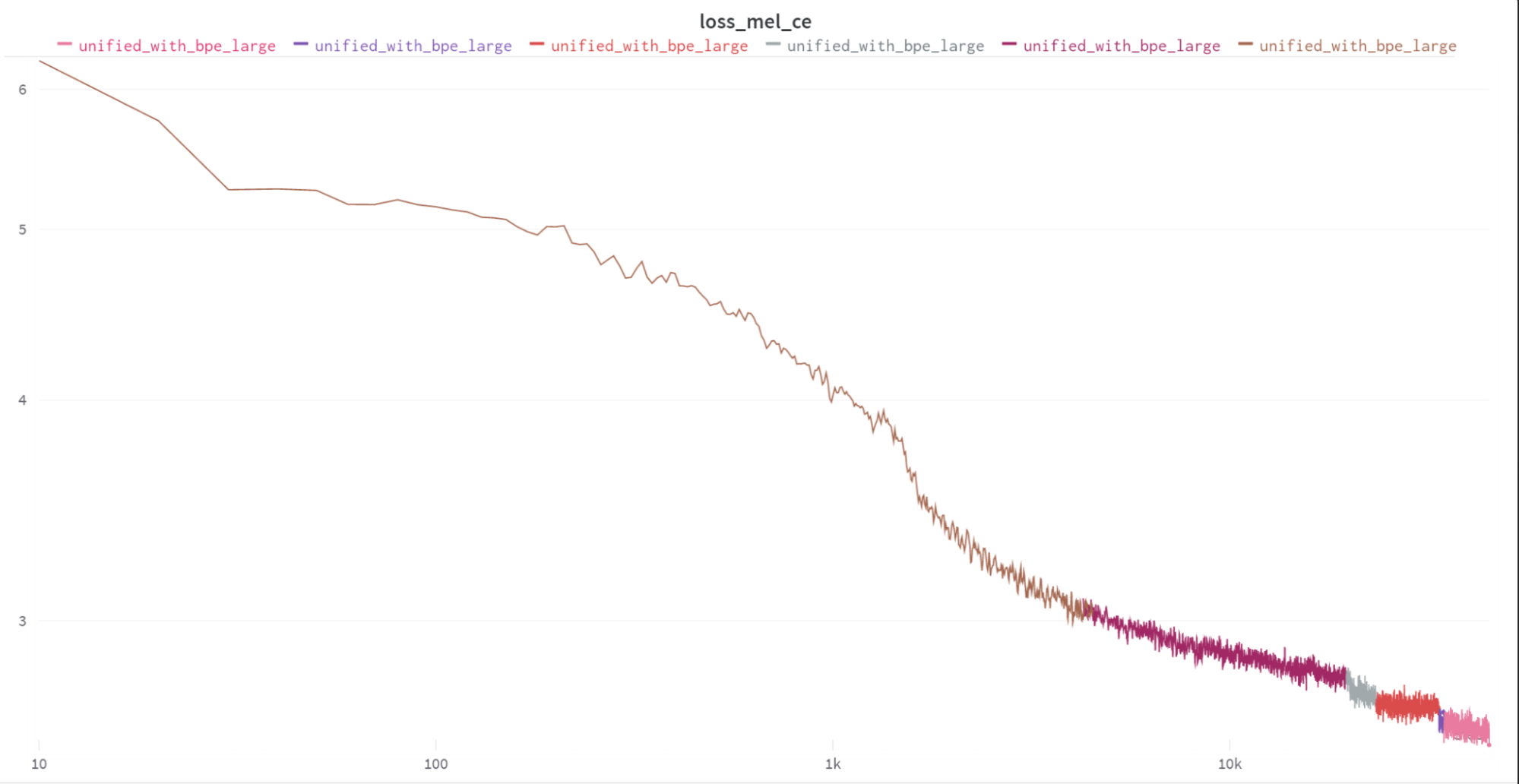}
    \end{center}
    \vskip -0.1in
    \caption{Early training curves in log-log scale. Y-axis is cross entropy loss for MEL tokens. X-axis is number of training steps. Does not include a long tail of training and fine-tuning due to online changes that were made adding non-reproducible noise to curves.}
    \vskip -0.1in
\end{figure}
\begin{table}[h]
    \setlength\tabcolsep{4pt}
    \begin{small}
    \begin{center}
    \begin{tabular}{ll}
    \toprule
        Model architecture & Transformer stack with causal masking \\
        Layers & 30 \\
        Model dim & 1024 \\
        Attention heads & 16 \\
        Text tokenization & Custom BPE, 256 tokens wide. \\
        Batch size & 1024 \\
        Total training & 119M samples \\
        Text, next token prediction, loss weight & .01 \\
        MEL token, next token prediction weight & 1 \\
        LR & 1e-4 \\
        B1, B2 & .9 .96 \\
        Weight decay & .01 \\
        LR Warmup & 500 steps \\
        EMA decay rate & .999 \\
    \bottomrule
    \end{tabular}
    \end{center}
    \end{small}
    \caption{AR prior details \& hyperparameters}
    \vskip -.2in
\end{table}

After training the autoregressive decoder to convergence, I fine-tuned it on the clean audio datasets from LibriTTS and HIFITTS.

\newpage
\subsection{CLVP}

The original DALLE worked by decoding a large number of images for a given text prompt, which were then fed through CLIP. The image that CLIP deemed closest to the input text was used as the final output.

I continue following this lead for TorToise, for reasons that will become evident in the results section. I built a simple model that is very similar to CLIP, which I call a “Contrastive Language-Voice Pretrained” model, or CLVP. Like CLIP, this model produces distance metrics for text/speech pairs. 

CLVP uses an architecture similar to the CLIP text encoder, except it uses two of them: one for text tokens and the other for MEL tokens. Tokens from both encoders were dropped out at a rate of 15\%. Fixed positional embeddings were used. Maximum text input length was 350 tokens (in practice never actually seen). Maximum MEL token input length was 293, or ~13 seconds of audio.

\begin{figure}[h]
    \begin{center}
    \includegraphics[width=\textwidth]{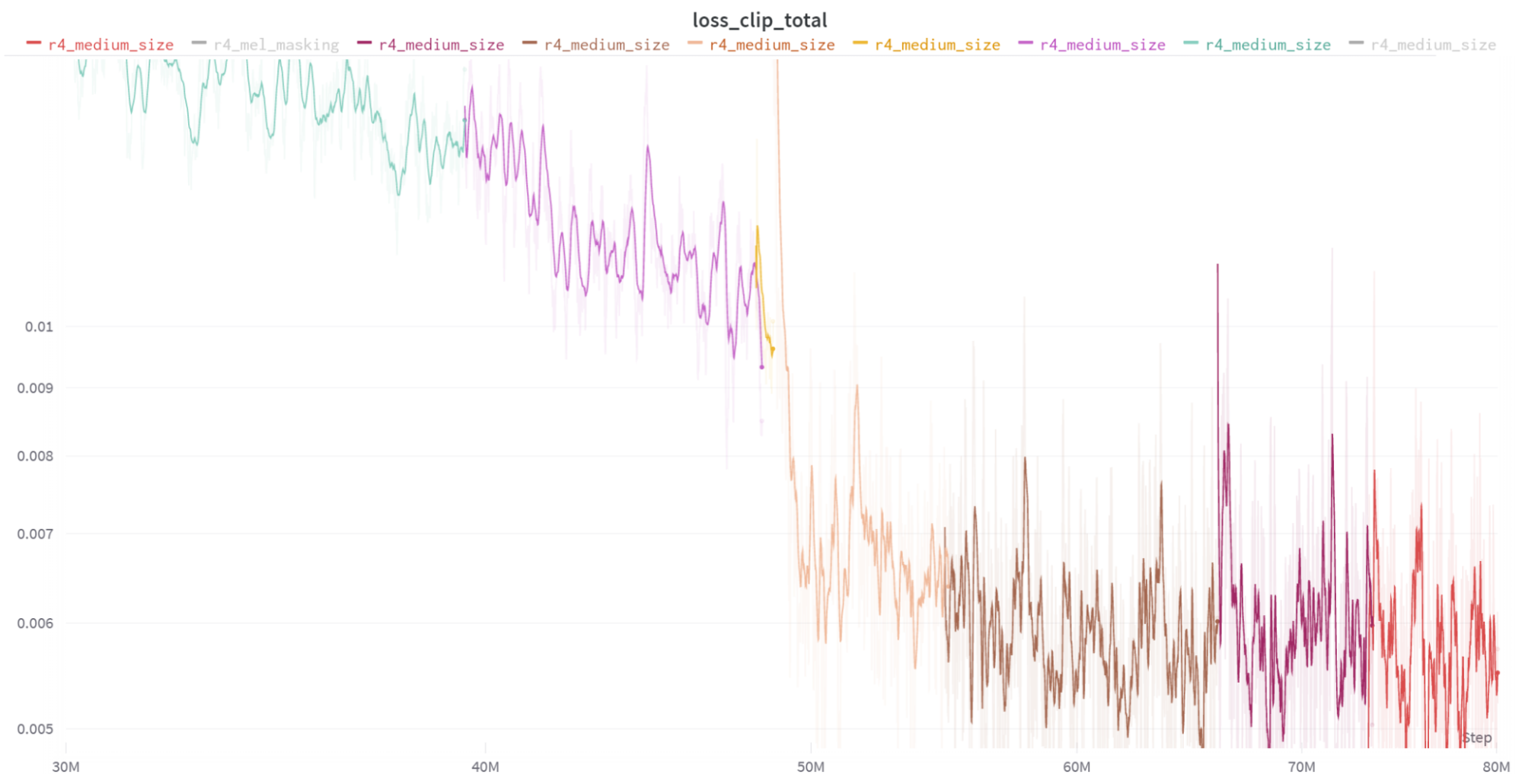}
    \end{center}
    \vskip -0.1in
    \caption{Late training curves for CLVP in log-log scale. Y-axis is cross entropy loss. X-axis is number of samples. Early training curves were lost.}
    \vskip -0.1in
\end{figure}
\begin{table}[h]
    \setlength\tabcolsep{4pt}
    \begin{small}
    \begin{center}
    \begin{tabular}{ll}
    \toprule
        Model architecture & Dual transformer stacks \\
        Depth & 20 \\
        Model dim & 768 \\
        Attention heads & 12 \\
        Text tokenization & Custom BPE, 256-token wide \\
        Batch size & 1024 \\
        Total training & 80M samples. \\
        Losses & Contrastive \\
        LR & 3e-4 \\
        B1, B2 & .9 .96 \\
        Weight decay & .001 \\
        LR Warmup & 500 steps \\
        EMA decay rate & .999 \\
    \bottomrule
    \end{tabular}
    \end{center}
    \end{small}
    \caption{CLVP training details \& hyperparameters}
    \vskip -.2in
\end{table}

\newpage
\subsection{Diffusion Decoder}
The diffusion model uses a bespoke architecture that combines residual convolutions with dense self-attention. It most closely resembles the traditional U-Net model used for DDPMs but without any upsampling or downsampling.

The diffusion model receives 3 sources of conditioning:
The timestep signal, which modulates the scale and shift of the group norms used by the network.
A speech conditioning signal, which also modulates the scale and shift of the group norms.
The final activations of the autoregressive model.

In training the diffusion model, I iterated through several different architectures and conditioning types before settling on this one. This includes:
Architecture: A “traditional” U-net with attention was tried. The full attention network performed significantly better in frechet distance evaluations.
Operating on PCM data rather than MELs. This necessitated very small context windows and still took an inordinate amount of time to train. The results of decoding a MEL and using a vocoder resulted in substantially better quality. In order to force compatibility with existing diffusion noise schedules, I rescale input MELs to be on the interval [-1,1].
Decoding MEL tokens versus AR activations. Training on AR activations is expensive because during each training step you must forward prop through the AR network. However, training on AR activations constituted the single greatest jump in output quality of any design decision made for the diffusion network. It is possible that doing tricks like putting the text on the attention context may ablate this advantage somewhat.

As with image diffusion models, exploiting classifier-free guidance is extremely important for high quality outputs. In the case of TorToise, I perform guidance on both the speech conditioning signal and the activations of the AR model. During training, 15\% of the time, both of these signals are dropped out and replaced with a learned embedding.

When training the diffusion decoder, input audio was clipped randomly to 220,500 samples, or 10 seconds of audio. Conditioning inputs were clipped to 102,400 samples, or ~5 seconds of audio. 

While the rest of the TorToise stack operates at an audio sampling rate of 22kHz, the diffusion decoder outputs MEL spectrograms which were computed from 24kHz audio. This discrepancy is solely to ensure compatibility with the pretrained Univnet vocoder which the model stack uses, and was not done for any performance reasons.

\begin{figure}[h]
    \begin{center}
    \includegraphics[width=\textwidth]{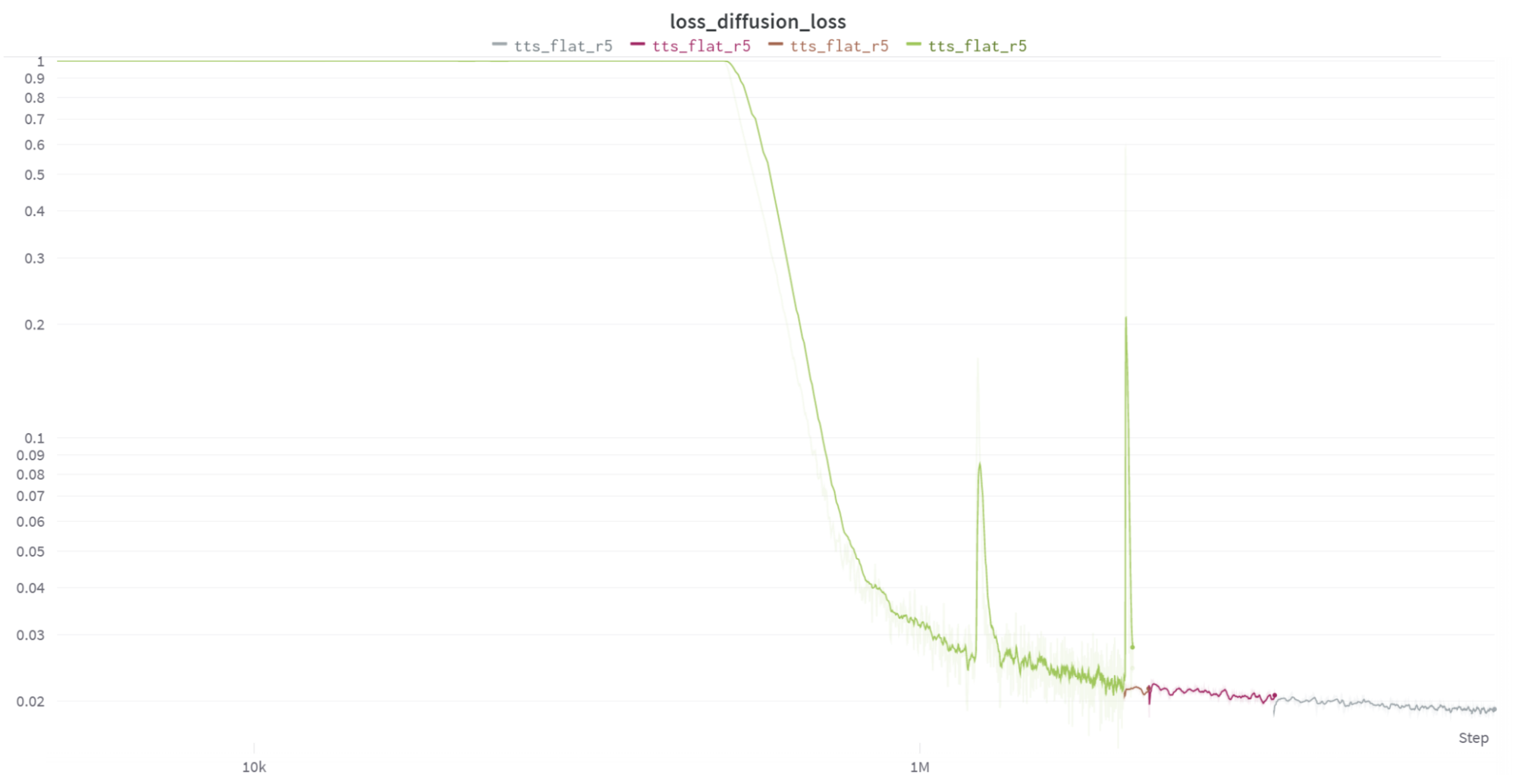}
    \end{center}
    \vskip -0.1in
    \caption{Diffusion model losses, log-log scale. Y-axis: MSE loss, X-axis: training samples.}
    \vskip -0.1in
\end{figure}
\begin{table}[h]
    \setlength\tabcolsep{4pt}
    \begin{small}
    \begin{center}
    \begin{tabular}{ll}
    \toprule
         Model shape & Alternating full attention + conv resblocks \\
         Depth & 10 \\
         Model dim & 1024 \\
         Attention heads & 16 \\
         Batch size & 512 \\
         Total Training & 65M samples \\
         Losses & MSE (weight 1) + VLB (weight n) \\
         LR & 1e-5 \\
         B1, B2 & .9, .999 \\
         Weight decay & .001 \\
         LR Warmup & 1000 steps \\
         EMA decay rate & .999 \\
    \bottomrule
    \end{tabular}
    \end{center}
    \end{small}
    \caption{Diffusion decoder details \& hyperparameters}
    \vskip -.2in
\end{table}

\newpage
\section{Future Work}
TorToise is the product of playing way over my paygrade, so to speak. As an independent researcher, I only had a small number of GPUs to perform my experiments with, and made many mistakes in the process. Following are recommendations for architectural tweaks to be made in future work building off of TorToise:
\begin{enumerate}
    \item Constrict VQVAE codebook embedding dim. This has been experimentally shown to produce drastic performance improvements.
    \item Relative positional encodings. The AR model uses fixed positional encodings, which limits the total amount of speech it can produce. Using relative encodings would allow arbitrary length sequences.
    \item Train CLVP on larger batch sizes. Contrastive models benefit from extremely large batch sizes.
    \item Train CLVP on longer audio sequences. CLVP only ever saw 13 second clips, which is likely why re-ranking on longer samples suffers.
    \item Diffusion decoder architecture. The diffusion decoder is an attentional network that omits Feedforward blocks. In retrosepct, this was a poor design decision and feed-forward blocks should be included.
    \item Train the entire model stack at 24kHz or re-train Univnet at 22kHz sampling rates.
    \item Train on more data for longer. The training curves for TorToise indicate that we were far from overfitting. Simply training longer likely would have improved results.
\end{enumerate}

\section{Special Thanks}

More than the prior work done by the research community, this project was a product of the open source community. I wanted to thank a few extra contributors who have not already been mentioned above whose work I found instrumental in building TorToiSe:

\begin{enumerate}
    \item Phil Wang, who authored \cite{lucidrainsx} and \cite{lucidrainsvq}.
    \item Kim Seonghyeon, who authored \cite{rosinalityvq}.
    \item FAIR, who maintain most of the tooling I use and who open sourced much of the technology underpinning TorToiSe.
    \item Prafulla Dhariwal and Alex Nichol, without whose \cite{improved_diffusion} I would still be in GAN hell.
\end{enumerate}

I also want to thank my wife, Kim Betker, who supported me through two years of high electricity bills, a hot \& noisy utility room, and the many late nights required to build this system.

\end{document}